\documentclass[letterpaper,conference,final]{IEEEtran}

\usepackage[binary-units=true]{siunitx}
\DeclareSIUnit \dBm {dBm}
\DeclareSIUnit \dB {dB} 
\DeclareSIUnit \dBi {dBi} 
\DeclareSIUnit \Kbps {Kbps}
\DeclareSIUnit \Mbps {Mbps}
\DeclareSIUnit \Gbps {Gbps}
\DeclareSIUnit \kBps {kBps}
\DeclareSIUnit \MBps {MBps}
\DeclareSIUnit \GBps {GBps}

\usepackage{cite}      
\usepackage[caption=false, font=footnotesize]{subfig}
\usepackage[pdftex]{graphicx}
\usepackage{url}       
\usepackage{amsfonts}
\usepackage{amssymb}
\usepackage{mathtools}
\usepackage{times}
\usepackage{algpseudocode}
\usepackage{fourier}
\usepackage{algorithm}
\usepackage{enumerate}
\usepackage{multirow}
\usepackage{tikz}
\usepackage{array}
\usepackage{gensymb}
\usepackage{eucal}
\usepackage{nicefrac}
\usepackage{tabularx}
\DeclarePairedDelimiter{\ceil}{\lceil}{\rceil}
\newcolumntype{P}[1]{>{\centering\arraybackslash}p{#1}}
\DeclareMathOperator*{\minimise}{\mathrm{Minimise}}
\bstctlcite{IEEEexample:BSTcontrol}
\newcommand*{\minimisel}{\minimise\limits}

\algnewcommand\algorithmicinput{\textbf{Phase 1:}}
\algnewcommand\Level{\item[\algorithmicinput]}

\algnewcommand\algorithmicinputt{\textbf{Phase 2:}}
\algnewcommand\Levell{\item[\algorithmicinputt]}

\algnewcommand\algorithmicinputtt{\textbf{Phase 3:}}
\algnewcommand\Levelll{\item[\algorithmicinputtt]}

\algnewcommand\algorithmicinputttt{\textbf{Output:}}
\algnewcommand\Output{\item[\algorithmicinputttt]}

\usepackage[english]{babel}
\usepackage{blindtext}

\begin{document}
\title{Efficient Millimeter-Wave Infrastructure Placement for City-Scale ITS}
\author{\IEEEauthorblockN{Ioannis Mavromatis\IEEEauthorrefmark{1}, Andrea Tassi\IEEEauthorrefmark{1}, Robert J. Piechocki\IEEEauthorrefmark{1}\IEEEauthorrefmark{3}, and Andrew Nix\IEEEauthorrefmark{1}}
  \IEEEauthorblockA{\IEEEauthorrefmark{1}Department of Electrical and Electronic Engineering, University of Bristol, UK \\
  \IEEEauthorblockA{\IEEEauthorrefmark{3}The Alan Turing Institute, London, NW1 2DB, UK}
  Emails: \{Ioan.Mavromatis, A.Tassi, R.J.Piechocki, Andy.Nix\}@bristol.ac.uk}
}

\maketitle

\begin{abstract}
Millimeter Waves (mmWaves) will play a pivotal role in the next-generation of Intelligent Transportation Systems (ITSs). However, in deep urban environments, sensitivity to blockages creates the need for more sophisticated network planning. In this paper, we present an agile strategy for deploying road-side nodes in a dense city scenario. In our system model, we consider strict Quality-of-Service (QoS) constraints (e.g. high throughput, low latency) that are typical of ITS applications. Our approach is scalable, insofar that takes into account the unique road and building shapes of each city, performing well for both regular and irregular city layouts. It allows us not only to achieve the required QoS constraints but it also provides up to $50\%$ reduction in the number of nodes required, compared to existing deployment solutions.
\end{abstract}

\section{Introduction}
Next-generation Intelligent Transportation Systems (ITSs) will require Vehicle-to-Infrastructure (V2I) wireless connectivity. These communication links offer the potential to enhance the road safety and efficiency in urban vehicular environments~\cite{necessityV2ILinks}. Introducing Millimeter-Waves (mmWaves) for dense urban environments will significantly improve the performance of small-cell access networks~\cite{mmWave5GCellular}. As shown, mmWave-V2I links have the potential of enabling gigabit-per-second data rates and ultra-low latency~\cite{vtcIoannis,tvtTassi}.

The communication capabilities of the vehicles are highly dependent on the number of deployed Road-Side Units (RSUs) and their coverage range. RSUs, are costly to deploy and maintain. Therefore, compromises between the coverage provided and the deployment costs have to be made. MmWave RSUs especially, are bounded by their Line-of-Sight (LOS) requirements and their strict propagation characteristics~\cite{losBoundMmWaves}. However, they are a perfect candidate for small-cell vehicular deployments as they can meet the rigid bitrate and latency Quality-of-Service (QoS) constraints needed by next-generation vehicular applications~\cite{vtcIoannis}.

It is crucial to deploy a number of RSUs in the most suitable locations, to improve the overall network performance. Given the variety of urban environments, it is necessary to find an agile method to obtain the best locations for each street layout and to deploy thousands of RSUs throughout a city. In this paper, we propose a strategy that can automate the RSU placement process for different urban scenarios. We will take into account the unique road and building layout of an urban environment as well as the strict QoS constraints of vehicular applications. Within a city, buildings may be away from a road or privately owned. Also, city blocks may be empty or covered with vegetation. On the other hand, traffic lights are usually placed at road intersections. Also, street furniture (e.g. lamp posts) are typically equally spread along the sides of a road. In this paper, we assume that our RSUs are deployed on top of lamp posts and traffic lights. Thus, by positioning the RSUs only on the road, easier access for deployment or maintenance is provided. Furthermore, the network efficiency could be improved, as the RSUs are more centrally located on the road, and thus avoiding the wall and rooftop blockages~\cite{streetFurniture}.


In~\cite{automatedMmWaveBSPlacment} and \cite{mmWaveComputationalGeometry}, the authors presented an automated base station placement algorithm for mmWaves. However, they did not consider any QoS constraints for their optimization algorithm apart from the LOS coverage rate. In our approach, we consider two main Key Performance Indicators (KPIs). The first one is the LOS network coverage achieved after the deployment of all the chosen RSUs. The second one is the Received Signal Strength (RSS) averaged throughout the considered deployment area. A similar work can be found in~\cite{ieee80211pRSUplacement}, where the authors proposed a strategy for placing IEEE 802.11p RSUs. As their optimization variable, they considered the delay tolerance of the warning notifications and solved their optimization problem utilizing a Genetic Algorithm (GA)~\cite{geneticAlgorithm}. However, mmWave links behave differently compared to IEEE 802.11p ones. In this paper, we will address the problem through a novel approach to find the best RSU locations by taking into account the propagation characteristics in mmWaves environments. Similarly, authors in ~\cite{rsuPlanningExample1,rsuPlanningExample2}, only considered the distance or the propagation characteristics, but not the shape of the buildings and the roads. In our case, we will utilize tools from Computational Geometry (as in~\cite{mmWaveComputationalGeometry}), and consider all the above to find the desirable RSU locations. More specifically, we will investigate the performance of our algorithm under two urban environments, describing their unique characteristics. Our outcome will be compared with solutions obtained by GA and Greedy Construction (GC)~\cite{greedyAddition} algorithms, to strengthen the enhanced performance provided by our approach.


The rest of the paper is organized as follows. In Sec.~\ref{sec:systemModel} we describe our system model. We introduce the tools utilized from Computational Geometry, and how we identify the candidate RSU positions. In Sec.~\ref{sec:problemSolution} we formulate our optimization problem based on two QoS constraints. Later, we outline our proposed strategy, solving the above problem. Sec.~\ref{sec:simResults} presents our performance investigation where we compare our strategy against the GC and GA approaches described before. Finally, Sec.~\ref{sec:conclusions} summarizes our findings.

\section{System Model}\label{sec:systemModel}
We consider an urban city map $\mathcal{M}$ with dimensions $\left[ \mathcal{M}_{x}, \mathcal{M}_{y} \right]$, measured in meters. Let $\mathcal{C} \triangleq \left \{1,\ldots,C \right \}$ denote the candidate RSU positions, with all being within the boundaries of $\mathcal{M}$. For all the above positions we denote as $\mathcal{D} \triangleq \left \{1,\ldots,D \right \}$ the positions chosen to deploy an RSU, and $\mathcal{R} \triangleq \left \{1,\ldots,R \right \}$ the rejected ones. We have that $\mathcal{D}\subseteq\mathcal{C}$, $\mathcal{R}\subseteq\mathcal{C}$, $\mathcal{D}\cup\mathcal{R}=\mathcal{C}$, and $\mathcal{D}\cap\mathcal{R}=\emptyset$ hold.

Our RSU placement algorithm works in two steps:
\begin{enumerate}
    \item At first, we identify all locations $\mathcal{C}$ to deploy an RSU employing tools from Computational Geometry.
    \item Then, we choose a subset $\mathcal{D}\subseteq\mathcal{C}$ in order to maximize the outdoor coverage and the RSS for the entire map.
\end{enumerate}

In our system model, we assume that all the RSUs are mounted at the height of a lamp post or a traffic light. We also assume that all vehicles' antenna will be mounted on their rooftop. By that, we avoid most of the low-level obstacles (for e.g., kiosks, vehicles, trees, etc.). For simplicity, we consider a 2D network planning. Finally, we make use of OpenStreetMap~\cite{OpenStreetMap} to obtain our city maps. 

\subsection{Identifying Potential RSU Locations}\label{sub:potentialLocations}
The problem of identifying $\mathcal{C}$ can be approached by using tools from Computational Geometry. Similarly to~\cite{mmWaveComputationalGeometry}, we introduce the notions of \emph{Simple Polygons (SPs)} and \emph{Polygons With Holes (PWHs)} in our system, describing the buildings and the roads, respectively. 

\begin{figure}[t]     
\centering
\includegraphics[width=1\columnwidth]{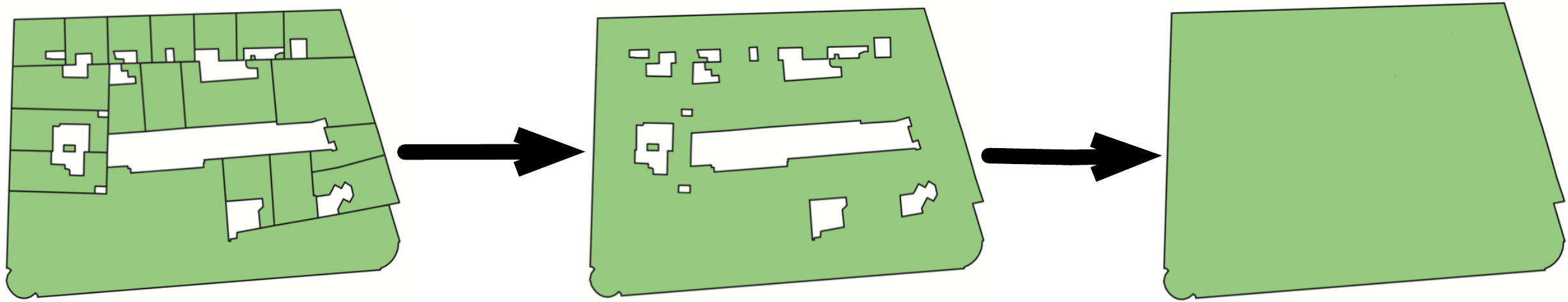}

    \caption{Example of the polygon union operation for a given city block.}
    \label{fig:unionPolygonOper}
\end{figure}

A SP is considered a flat-shaped object consisting of straight, non-intersecting line segments, that, when joined pair-wise, they form a closed path. Given a city map, many SPs have adjacent sides, being part of the same city block. Therefore, they can be concatenated using the polygon union operation~\cite{polygonUnion}. Furthemore, concatenating all the adjacent SPs, we may end up having a hole in the middle (e.g. a courtyard). These inner holes can be later removed forming a solid object (as in Fig.~\ref{fig:unionPolygonOper}). By that, and based on the nature of mmWaves (the signal intersecting with an object will result in a blockage), we can decrease the complexity of our algorithm and, consequently, the execution time without any loss of accuracy. Furthermore, PWH is a polygon with an irregular shape that contains one or more holes or cutouts in it. Having access to the metadata from OpenStreetMap, we can accurately calculate the different polygons representing each road. Each polygon is later concatenated with the others, having finally a concave PWH that will be used for the RSU placement. The SPs introduced before, will be used to determine if a link is in LOS or not.

Given the generated SPs and PWHs, we can identify $\mathcal{C}$ for a particular map. Our algorithm searches along the sides of PWHs for sharp edges and long straight sections. The edges, being the corner of two roads, are usually the best positions for an RSU, as they can maximize the LOS coverage (as shown in~\cite{automatedMmWaveBSPlacment}). We also consider the length of a road. A road qualifies as a ``long road'' when the distance between two intersections is greater than a given threshold $\mathrm{RSU}_{t}$. For any long road, we consider more potential positions, equally spaced between the two intersections. The number of these positions is given as the ceiling function from the division of the length $l_i$ of the road i, over the given threshold, i.e., $\ceil{\nicefrac{l_i}{\mathrm{RSU}_{t}}}$. Combining both lists, $\mathcal{C}$ is found.

\section{Problem Formulation and Solution}\label{sec:problemSolution}
Given $\mathcal{C}$, the objective is to find $\mathcal{D}\subseteq\mathcal{C}$ to maximize the network coverage and achieve a minimum RSS throughout the network. The LOS coverage rate is modeled by equally spacing $\mathcal{Z} \triangleq \left \{1,\ldots,Z \right \}$ grid points on the map with equal weights. Each point represents a squared tile having the same RSS throughout its surface. Using a tile-like approach with relatively small tiles, we can decrease the processing power required without having a significant loss of accuracy. 

We determine $\mathcal{D}$ by taking into account two different constraints. For a given $\mathcal{Z}$, we consider a set of $\mathcal{N}$ reference points, with $\mathcal{N}\subseteq\mathcal{Z}$, being the grid points on top of the road polygons. We define the binary variable $\gamma_n$, to denote the state of a reference point $n$ at location $(x,y)$ as follows:
\begin{equation}
\gamma_{n}(x,y) = 
  \begin{cases}
      1,~\mathrm{if}~n~\mathrm{is~covered~by~at~least~one~RSU}\\
     0,~\mathrm{otherwise}
  \end{cases}
\end{equation}
Our first constraint imposes that a number of tiles, subset of $\left\vert \mathcal{N} \right\vert$, are in LOS with at least one RSU, i.e.:
\begin{equation}\label{eq:constraint1}
\sum\limits_{n=1}^{\mathcal{\left\vert N \right\vert}}\gamma_{n}(x,y) \geq \tau\,\left\vert \mathcal{N} \right\vert
\end{equation}
where $\tau \in \left[ 0,1 \right]$, is a tolerance factor.

We say that $P_{\mathrm{tx}}$ and $G_{\mathrm{tx}}$ are the transmission power and antenna gain of each RSU. Also, $L_{\mathrm{LOS}}(d)$ is regarded as the propagation loss at a distance $d$. Given the above, we can calculate the RSS for each tile as:
\begin{equation}
\mathrm{RSS}(d) = P_{\mathrm{tx}} + G_{\mathrm{tx}} - L_{\mathrm{LOS}}(d)
\end{equation}
where $L_{\mathrm{LOS}}(d)$ is the path-loss component and can be calculated as $L_{\mathrm{LOS}}(d) = 10 \, \alpha \, \log_{10}(d) + C_{\mathrm{att}}(d)$.
$C_{\mathrm{att}}(d)$ is the channel attenuation with regard to the distance $d$. It is defined by the rain and atmospheric attenuation as well as the channel attenuation factor $H_{\mathrm{att}}$ for a given mmWave LOS link at \SI{60}{\giga\hertz} in urban environments~\cite{prediction_model}, i.e. $C_{\mathrm{att}}(d) = 40d + H_{\mathrm{att}}$. Finally, $\alpha$ is the path loss exponent.

For all $\mathcal{D}$, there is a number of tiles $K_i$ that surrounds it. Each tile can be served by more than one RSU. We define as $\mathrm{max}_{i \in \left\lbrace 1,\ldots,\tau\,\left\vert \mathcal{N} \right\vert \right\rbrace} \, \{K_i^\mathrm{RSS}\}$ the highest received RSS from all RSUs that serve this tile. The interference between the deployed RSUs, is not taken into account. For the entire covered area, we sort all tiles with respect to their received RSS value, we take the first $\tau\, \left\vert \mathcal{N} \right\vert$ and denote them as $\mathrm{RSS_{i,k}^{\mathrm{max}}}$. $\mathrm{RSS_{i,k}^{\mathrm{max}}}$ is the average for the best $\tau\, \left\vert \mathcal{N} \right\vert$ tiles and for a given $\mathcal{D}_i$
Our second constraint ensures that a target number of tiles has an average RSS that is greater than or equal to a threshold $\mathrm{RSS}_{th}$:
\begin{equation}
    \sum\limits_{i=1}^{\left\vert \mathcal{D}\right\vert} 
    \,
    \mathrm{RSS_{i,k}^{\mathrm{max}}}
    \, \geq \, \mathrm{RSS}_{th}
\end{equation}

\subsection{Problem Formulation}
Let $e_\mathcal{C}$ be the vector that defines the state of each RSU $i$ in $\mathcal{C}$. We have:
\begin{equation}\label{eq:fitness}
e_{i} = 
  \begin{cases}
      1,~\mathrm{if}~i~\mathrm{is~deployed}\\
     0,~\mathrm{if}~i~\mathrm{is~not~deployed}
  \end{cases}
\end{equation}

\begin{figure}[t]     
\centering
\includegraphics[width=\columnwidth]{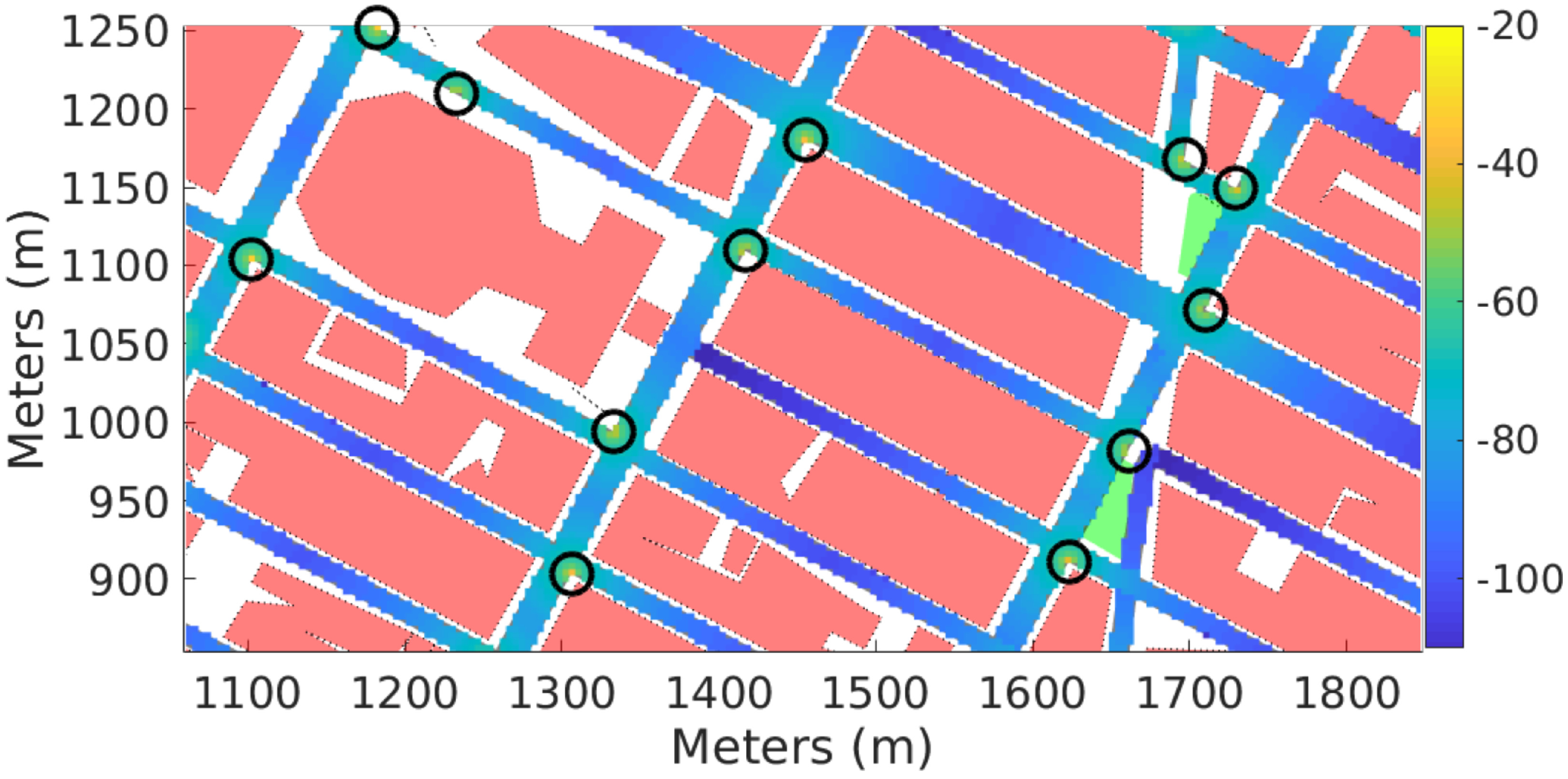}

    \caption{Example of 12 RSUs chosen for Manhattan by our strategy with their corresponding RSS. The street geometry affects the perceived RSS.}
    \label{fig:rsuPlacement}
\end{figure}

In order to find the best RSU locations, our problem is formulated as the minimization of \eqref{eq:fitness}:
\begin{subequations}\label{eq:problem}
\begin{align}
\displaystyle \minimisel_{e_1,\ldots,e_\mathcal{C}} & \displaystyle\sum\limits_{i=1}^{\left\vert \mathcal{C} \right\vert}e_{i} & \label{eq:5a}\\ 
\textrm{subject to:} & \displaystyle \sum\limits_{n=1}^{\mathcal{\left\vert N \right\vert}}\gamma_{n}(x,y) \geq \tau\,\left\vert \mathcal{N} \right\vert &\label{eq:5b} \\
& \sum\limits_{i=1}^{\left\vert \mathcal{D}\right\vert} \,  \mathrm{RSS_{i,k}^{\mathrm{max}}} \, \geq \, \mathrm{RSS}_{th} & , \forall i \in \mathcal{D} \label{eq:5c} \\
 & e_{i} \in \left[ 0,1 \right] & , \forall i \in \mathcal{C} \label{eq:5d}  \\
 & \mathrm{RSS_{i,k}^{\mathrm{max}}}~\mathrm{for~}k~\mathrm{in~LOS} & , \forall i \in \mathcal{D} \label{eq:5e}
\end{align}
\end{subequations}
where the above are the constraints described before. Also, the RSS is calculated for all tiles being in LOS with an RSU, i.e. a SP does not block the link. An example solution for a small urban area can be seen in Fig.~\ref{fig:rsuPlacement}. In this particular example, we can observe the effect of the street geometry on the RSS.

\subsection{Proposed Algorithm}
To solve~\eqref{eq:5a}-\eqref{eq:5e}, we propose a novel algorithm to calculate the list $\mathcal{D}$. A city-scale RSU placement problem can be computationally expensive. So, we designed our algorithm to operate in three phases. This can minimize the execution time required for our final solution. Our algorithm (Alg.~\ref{alg:autoPlacement}) works are as follows:
\subsubsection{Phase 1} 
We start by calculating the number of tiles $k_{i}^{'}$ required to achieve a mean RSS value greater than or equal to $\mathrm{RSS}_{th}$ for all $\mathcal{C}$. Later, we iteratively add to $\mathcal{D}$ the RSU with the most non-served tiles within $\mathrm{RSS}_{th}$ found before, until constraint~\eqref{eq:5b} is met. Thus, we ensure a sufficient amount of coverage in our system in a fast greedy-addition-like fashion. If~\eqref{eq:5c} is also fulfilled, we proceed to Phase 3. If not, we continue with Phase 2.
\subsubsection{Phase 2} 
We add more RSUs in the system until both constraints are met. We identify the non-covered areas of the map and prioritize our RSU placement towards them, as this can increase our system performance. Then, we find the areas that are not adequately served by the existing RSUs, and we add RSUs that can fulfill~\eqref{eq:5c}. When both constraints are fulfilled, we have two admissible lists $\mathcal{D}$ and $\mathcal{R}$ and we proceed to the next phase.
\subsubsection{Phase 3} From the above two phases, we may not always achieve an ideal solution for $\mathcal{D}$. This happens especially when the requirements for a specific scenario are more relaxed (e.g. a low coverage rate is required). To improve the performance at this point, we search in $\mathcal{R}$, if it exists an RSU that can improve~\eqref{eq:5b} or~\eqref{eq:5c}. If so, we replace these two. We iterate, until no other RSUs can be swapped, meaning that we have our final $\mathcal{D}$.

\begin{algorithm}[t]

\caption{Agile RSU Placement}
\label{alg:autoPlacement}
{\footnotesize\begin{algorithmic}[1]
    \Output Return lists with RSUs: $\mathcal{D}$ and $\mathcal{R}$
    \Level
    \State Calculate the tiles $k_{i}^{'}$ required to achieve the $\mathrm{RSS}_{th}$ for each $\mathcal{C}$
	\While{Constraint~\eqref{eq:5b}~not~met}
	    \State Find served tiles in the system and remove from lists $k_{i}^{'}$.
	    \State Find RSU $i$ with the longest $k_{i}^{'}$ list and add it in $\mathcal{D}$.
    \EndWhile
    \Levell -- Skip if~\eqref{eq:5b} and~\eqref{eq:5c} are met.
	\While{Constraint~\eqref{eq:5c} is not met}
	    \ForAll{RSUs in $\mathcal{C} \not\in \mathcal{D}$}
	        \State Calculate number of non-covered tiles that they can serve.
	    \EndFor 
	    \If{Non-covered tiles on map}\Comment{i.e., \eqref{eq:5b} is not maximised}
	        \State Find RSU $i$ that covers the most non-covered tiles.
	        \State Add $i$ in list of chosen RSUs $\mathcal{D}$.
        \Else
            \ForAll{RSUs in $\mathcal{C} \not\in \mathcal{D}$}
	            \State Calculate the potential mean RSS if RSU is chosen.
	        \EndFor
	        \State Find RSU $i$ that maximises the mean RSS for the system.
	        \State Add $i$ in list of chosen RSUs $\mathcal{D}$.
        \EndIf
    \EndWhile
    \Levelll
    \Repeat
        \ForAll{RSUs $i$ in $\mathcal{D}$}
	        \State Find $k$ in $\mathcal{C} \not\in \mathcal{D}$ that improves constraints~\eqref{eq:5b} and~\eqref{eq:5c}.
	        \State Replace $i$ with $k$.
	    \EndFor
    \Until{$\mathcal{D}$ cannot be improved more} \Comment{i.e., no more swaps can be done}
\end{algorithmic}}
\end{algorithm}

\section{Simulation Results}\label{sec:simResults}

\begin{table}[t]
\renewcommand{\arraystretch}{1.07}
\centering
\caption{List of Map Areas Used and Simulation Parameters.}
\begin{tabularx}{\columnwidth}{*{1}{p{.28\columnwidth}}*{2}{P{.30\columnwidth}}}
\raggedleft\textbf{Urban Area} & \textbf{Manhattan, NY, USA}  & \textbf{Paris, FR}  \\ \hline \hline
\raggedleft Centre & $-73.9841\degree W, 40.75545\degree N$   &  $2.33235\degree W, 48.875\degree N$ \\
\raggedleft Number of Maps & $2 \times 2$ & $2 \times 2$  \\
\raggedleft Map Size $\left[ \mathcal{M}_{x}, \mathcal{M}_{y} \right]$ & \SI{1}{\kilo\meter} $\times$ \SI{1}{\kilo\meter} & \SI{1}{\kilo\meter} $\times$ \SI{1}{\kilo\meter} \\
\raggedleft Area of Interest & \SI{900}{\meter} $\times$ \SI{900}{\meter}  & \SI{900}{\meter} $\times$ \SI{900}{\meter}  \\
\raggedleft Grid-Tile Size & $\SI{4}{\meter} \times \SI{4}{\meter}$ & $\SI{4}{\meter} \times \SI{4}{\meter}$ \\ \hline
\end{tabularx}
\newline
\vspace*{0.3cm}
\newline
\begin{tabularx}{\columnwidth}{*{1}{p{.28\columnwidth}}*{1}{P{.30\columnwidth}}*{1}{P{.32\columnwidth}}}
\raggedleft\textbf{Parameter} & \textbf{Symbol}  & \textbf{Value}  \\ \hline \hline
\raggedleft Transmission power  & $P_{\mathrm{tx}}$  & \SI{10}{\dBm}  \\
\raggedleft TX Antenna Gain     & $G_{\mathrm{tx}}$  & \SI{15}{\dBi} \\
\raggedleft Path-Loss Exponent  & $\alpha$       & 2.66~\cite{path_loss}    \\ 
\raggedleft Channel Att. Factor & $H_{\mathrm{att}}$  & \SI{70}{\dB}~\cite{prediction_model}  \\
\raggedleft Distance Threshold  & $\mathrm{RSU}_{t}$ & \SI{100}{\meter} \\ \hline
\label{tab:simParameters}
\end{tabularx}
\end{table}

\begin{figure*}[t]
\minipage{0.49\textwidth}
\centering
\includegraphics[width=1\columnwidth]{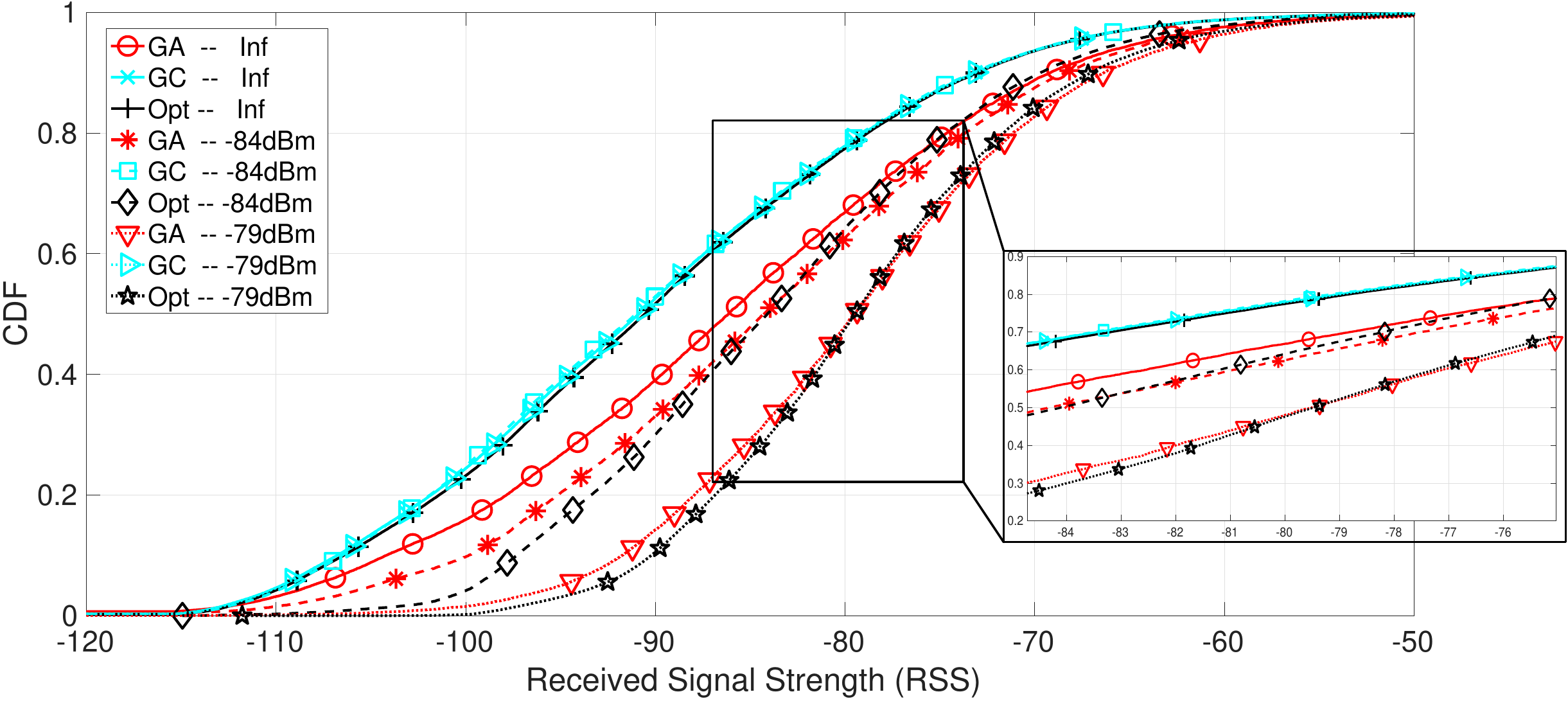}
    \caption{The empirical CDF of the RSS for the map of Manhattan. A tolerance $\tau=0.99$ and three different $\mathrm{RSS}_{th}$ were used for this scenario.}
    \label{fig:cdfManhattan}
\endminipage\hfill
\minipage{0.49\textwidth}
\centering
\includegraphics[width=1\columnwidth]{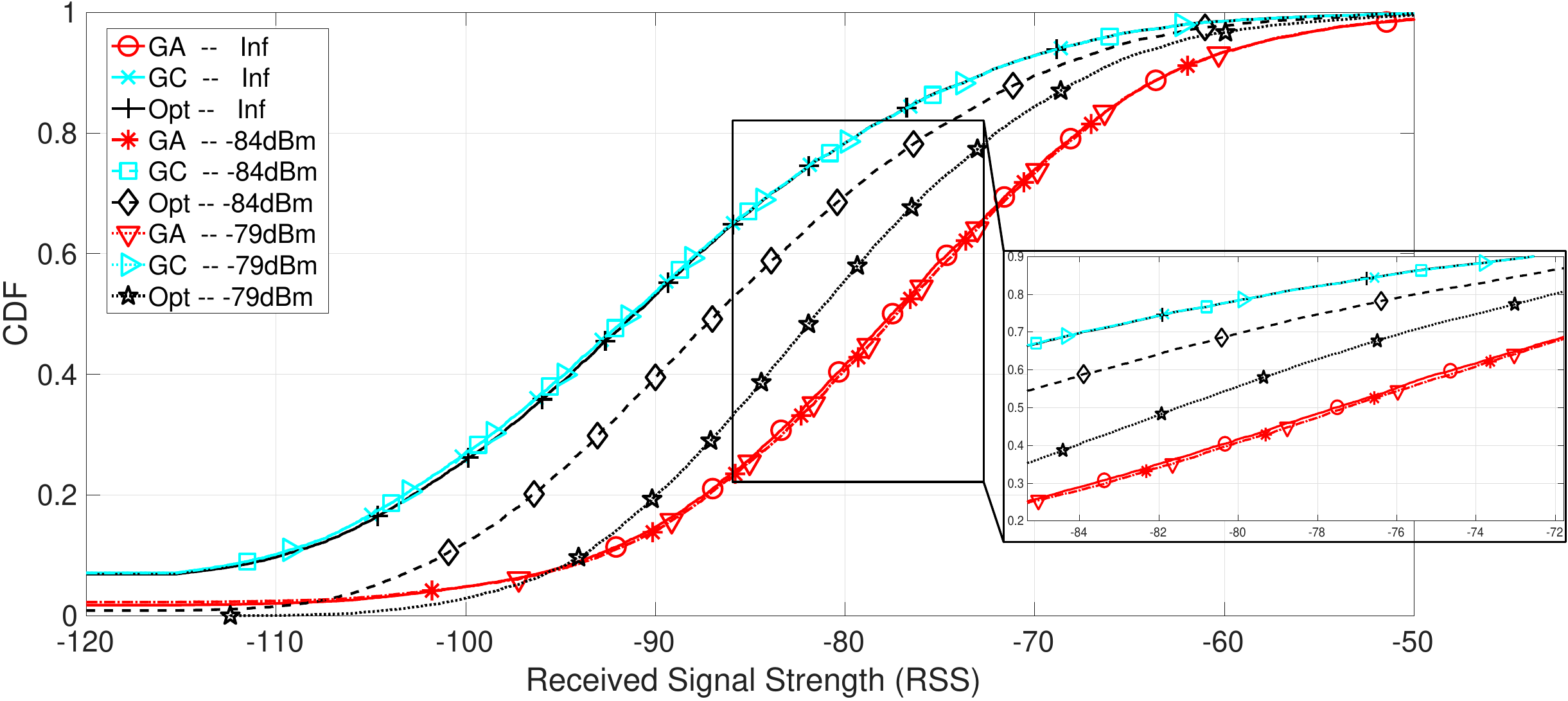}
    \caption{The empirical CDF of the RSS for the map of Paris. A tolerance $\tau=0.90$ and three different $\mathrm{RSS}_{th}$ were used for this scenario.}
    \label{fig:cdfParis}
\endminipage\hfill
\minipage{0.49\textwidth}
\vspace*{0.4cm}
\centering
\includegraphics[width=1\columnwidth]{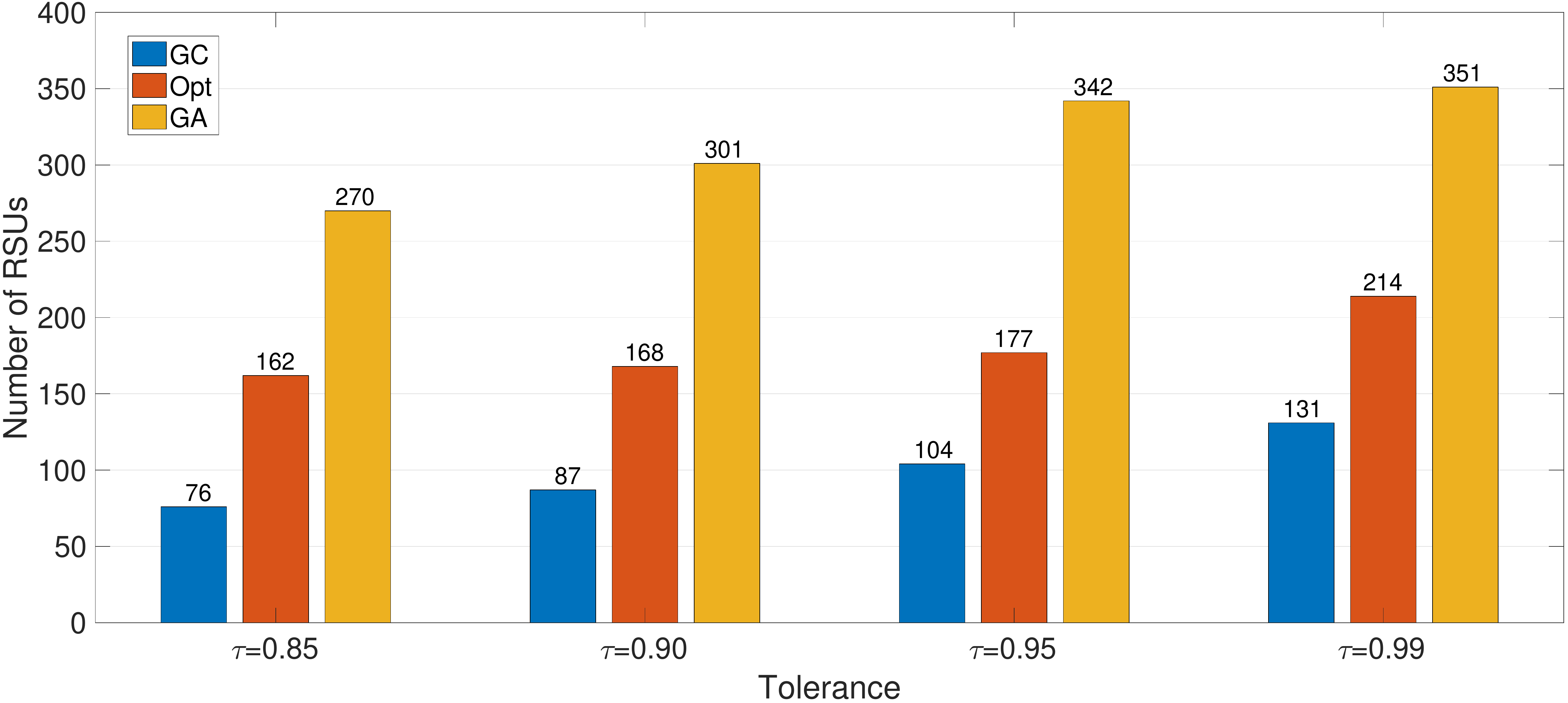}
    \caption{The number of RSUs (Manhattan), given by all algorithms, for all different tolerance parameters. The $\mathrm{RSS}_{th}$ is equal to \SI{-84}{\dBm}.}
    \label{fig:barManhattan}
\endminipage\hfill
\minipage{0.49\textwidth}
\centering
\includegraphics[width=1\columnwidth]{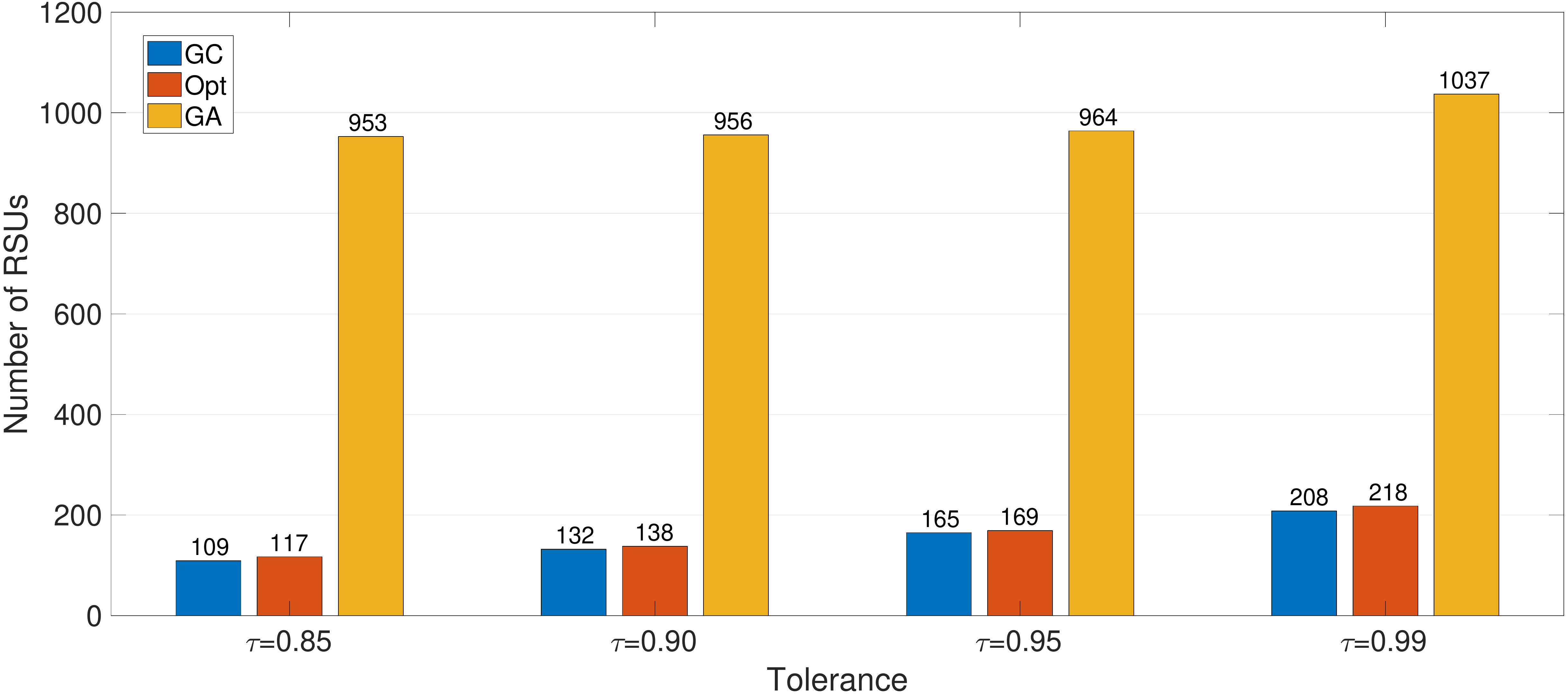}
    \caption{The number of RSUs (city of Paris), given by all algorithms, for all different tolerance parameters. The $\mathrm{RSS}_{th}$ is equal to \SI{-90}{\dBm}.}
    \label{fig:barParis}
\endminipage\hfill
\end{figure*}

Our strategy is compared against a GC and a GA approach. For GC, during each iteration, we choose an RSU to add in $\mathcal{C}$, finding the one that has the most LOS tiles from $\mathcal{N}$. We iterate until we meet~\eqref{eq:constraint1}. GC is scalable, but it cannot fulfill the required KPIs, as it does not take into account the RSS threshold. On the other hand,  GA is notoriously computationally expensive but generates high-quality solutions. We test all the algorithms in two urban areas from Manhattan (New York, USA) and Paris (FR). An average road lane is \SIrange{2.9}{4.6}{\meter} wide. Therefore, we considered a grid with side equal to \SI{4}{\meter}, so each tile covers roughly the width of a road lane. The considered areas and the simulation parameters are summarized in Table~\ref{tab:simParameters}. 

Each area is $\SI{4}{\kilo\meter}^2$ and divided into four equal sections. Each section is considered as an independent map in our simulation with dimensions $\left[ \mathcal{M}_{x}, \mathcal{M}_{y} \right]$ and a surface of $\SI{1}{\kilo\meter}^2$. The centre coordinates given in Table~\ref{tab:simParameters} present the point that the edges of all four sections meet. For each section, we consider an area of interest of $\SI{810}{\meter}^2$, to avoid border effects. Four different tolerance parameters are employed, namely, $\tau \in \left\lbrace 0.85, 0.90, 0.95, 0.99 \right\rbrace$), with 90\% being an average coverage rate, while 99\% being the extreme case. Also, four RSS thresholds are used, that is $\mathrm{RSS}_{th} \in \left\lbrace \mathrm{Inf}, -90, -84, -79 \right\rbrace$. The first value signifies the case where no RSS threshold is considered. The last one was chosen based on the sensitivity threshold of IEEE 802.11ad, i.e., the minimum RSS (without considering the RX antenna gain) to achieve one-gigabit-per-second data rate. This value is the amount of data estimated to be generated and transmitted from each Connected and Autonomous Vehicle (CAV)~\cite{qosReq}.

In Fig.~\ref{fig:cdfManhattan}, we present the RSS results per-tile for Manhattan and all the considered optimization strategies. Also, $\tau = 0.99$ and three different thresholds were considered, i.e. $\mathrm{RSS}_{th} \in \left\lbrace \mathrm{Inf}, -84, -79 \right\rbrace$, to investigate the effect on each approach. We observe that GC always produces similar results, comparable with our scheme (when the RSS constraint is disregarded). That is because GC cannot take into account the RSS threshold. Obviously, as the $\mathrm{RSS}_{th}$ is taken into consideration, we observe that both our algorithm and GA perform better than GC. In particular, our strategy achieved a city-wide mean RSS of \SI{-83.6}{\dBm} and \SI{-78.9}{\dBm}, while GA achieved \SI{-84.3}{\dBm} and \SI{-79.2}{\dBm} for $\mathrm{RSS}_{th} = \SI{-84}{\dBm}$ and $\mathrm{RSS}_{th} = \SI{-79}{\dBm}$, respectively. 

Similarly, Fig.~\ref{fig:cdfParis} presents the RSS results for Paris and $\tau = 0.90$. Again, GC algorithm behaves as before, having similar performance for all thresholds and being comparable with our algorithm when $\mathrm{RSS}_{th}$ is not taken into account. However, the main difference compared to the Manhattan results is the performance dissimilarity between our algorithm and GA. We observed that, even with relaxed parameters GA always finds an extreme solution (mean RSS of \SI{-78.8}{\dBm} and \SI{-78.5}{\dBm} for $\mathrm{RSS}_{th} = \SI{-84}{\dBm}$ and $\mathrm{RSS}_{th} = \SI{-79}{\dBm}$ respectively), while our algorithm behaves as before (mean RSS of \SI{-86.2}{\dBm} and \SI{-80.8}{\dBm}). This is because of the highly irregular building shapes of Paris, compared to the grid-like shape of Manhattan, making it very difficult for GA to find the best RSU positions in the city and always getting stuck to a local maximum. For both cities, all the algorithms behave similarly to other tolerance parameters and will not be presented here due to the limited space.

In Figs.~\ref{fig:barManhattan} and~\ref{fig:barParis}, we present the number of RSUs required to fulfill the QoS constraints required, for all $\tau$. Fig.~\ref{fig:barManhattan} refers to the Manhattan scenario, for $\mathrm{RSS}_{th} = \SI{-84}{\dBm}$. Once more, GC utilizes fewer RSUs, but it does not take into account the $\mathrm{RSS}_{th}$. Comparing the number of RSUs obtained with the proposed approach and with GA, we observe that our scheme ensures a reduction of up to $50$\% in the number of RSUs compared to GA. From Figs.~\ref{fig:cdfManhattan} and~\ref{fig:barManhattan}, we observe that our algorithm achieves comparable results, fulfilling the QoS requirements with a smaller number of RSUs. Moving on to Fig.~\ref{fig:barParis}, the difference between the number of RSUs and the two algorithms is greater, having GA solving our optimization problem with almost eight times as many RSUs. Again, as before, this is because of the irregularity of the building and road shapes. 

\section{Conclusions}\label{sec:conclusions}
In this paper, we proposed an agile and efficient strategy for city-wide mmWave-RSU placement. We presented a scalable algorithm, able to compute high-quality RSU deployment on a map. In doing so, the proposed strategy takes into account two KPIs for vehicular communications: the coverage rate and the RSS threshold. Our approach is compared against the GC and GA strategies. GC is fast and scalable, but it cannot satisfy the above mentioned KPIs, while GA is computationally expensive. We observed that our strategy meets the target coverage constraints utilizing a smaller number of RSUs. Also, our performance investigation showed that our approach is suitable for both regular and irregular city layouts when other strategies fail to handle non-uniform building shapes. All the above make it a suitable solution for large-scale experimentation and the next-generation ITSs.

\section*{Acknowledgment}
This work was partially supported by the University of Bristol and the Engineering and Physical Sciences Research Council (EPSRC) (grant ref. EP/I028153/1). This work is also part of the FLOURISH Project, which is supported by Innovate UK, under Grant 102582.

\bibliographystyle{IEEEtran}
\bibliography{bib.bib,IEEEabrv}

\end{document}